\documentclass[final,5p,twocolumn]{elsarticle}

\usepackage{amssymb}
\usepackage{hyperref}

\biboptions{sort&compress}

\begin{document}

\begin{frontmatter}

\title{Nonperturbative partonic quasidistributions of the pion from chiral quark models}

\author[ifj,ujk]{Wojciech Broniowski}
\ead{Wojciech.Broniowski@ifj.edu.pl}

\author[ugr]{Enrique Ruiz Arriola}
\ead{earriola@ugr.es}

\address[ifj]{The H. Niewodnicza\'nski Institute of Nuclear Physics, Polish Academy of Sciences, 31-342 Krak\'ow, Poland}
\address[ujk]{Institute of Physics, Jan Kochanowski University, 25-406 Kielce, Poland}
\address[ugr]{Departamento de F\'isica At\'omica, Molecular y Nuclear and Instituto Carlos I de
  Fisica Te\'orica y Computacional,  \\ Universidad de Granada, E-18071 Granada, Spain}

\date{ver. 1: 29 July 2017, ver. 2: 31 August 2017}

\begin{abstract}
We evaluate nonperturbatively the quark quasidistribution amplitude
and the valence quark quasidistribution function of the pion in the
framework of chiral quark models, namely the Nambu--Jona-Lasinio
model and the Spectral Quark Model. We arrive at simple analytic
expressions, where the nonperturbative dependence on the longitudinal
momentum of the pion can be explicitly assessed.
The model results for the quark quasidistribution amplitude of the pion compare 
favorably to the data
obtained from the Euclidean lattice simulations. The quark distribution amplitude, arising in the limit
of infinite longitudinal momentum of the pion, agrees, after suitable QCD evolution, to the recent 
data extracted from Euclidean lattices, as well as to the old data from transverse lattice simulations. 
\end{abstract}


\end{frontmatter}

\section{Introduction}

Partonic distributions provide direct access to the internal structure
of hadrons in terms of the Bjorken variable, $x$, but their {\it ab
initio} determination directly from QCD has remained elusive during
many years, with the only exception of isolated and courageous efforts
on the transverse lattice formulated {\it directly} on the
light cone~\cite{Burkardt:2001mf,Dalley:2002nj} (for a review see,
e.g., \cite{Burkardt:2001jg}).  There, both the parton distribution
functions (PDF) and the parton distribution amplitudes (PDA) of the pion
have been determined. The case of the pion, which is of our concern
here, is particularly interesting, since not only it is the lightest
hadron but also its properties are strongly constrained by the chiral
symmetry and its spontaneous breakdown, a feature difficult
to implement on the light-cone. In addition, the gauge invariance,
relativity, and positivity are essential ingredients in the calculation;
they guarantee correct normalization, proper support of the parton
distributions, and a probabilistic interpretation. However, for the
most popular implementation of QCD on Euclidean lattices only moments of
distributions in $x$ can be accessed as matrix elements of local
operators. The breaking of the Lorentz invariance of the lattice generates
operator mixing with the result that up to now only a few lowest moments
can be extracted with an acceptable signal to noise ratio. 

Within this state of affairs, Ji has proposed a scheme where
nonperturbative partonic physics could be accessed on a Euclidean
lattice by boosting the space-like correlators to a finite momentum
and, eventually, extrapolate the results to the infinite-momentum
frame~\cite{Ji:2013dva}. The resulting objects, termed parton
quasidistributions, may offer a unique missing link between
nonperturbative physics and the light-cone dynamics; their peculiar
properties have received a great deal of attention both from a
theoretical as well as from a practical perspective. Following this
proposal, in the non-singlet case, the one-loop matching for parton
distributions~\cite{Xiong:2013bka}, the renormalization of parton
quasidistributions~\cite{Ji:2015jwa}, as well as the one-loop matching
for generalized parton distributions~\cite{Ji:2015qla} have been
studied. Power divergences, ubiquitous in lattice QCD, can be safely
removed by introducing an improved parton quasidistribution through
the Wilson line renormalization~\cite{Chen:2016fxx}. A quite appealing
feature is the possibility of reformulating the whole problem as an
effective large-momentum effective theory approach to parton physics,
where the separation scale is the boosting momentum
$P_z$~\cite{Ji:2014gla,Ji:2017rah,Ishikawa:2016znu,Chen:2017mzz}.

From a lattice perspective, the
pion has always been a challenge, since realistic calculations require
very small quark masses which need very large lattice volumes.  The
most impressive implication of the mentioned series of works is the
recent determination of PDA of the pion on the Euclidean 
lattice~\cite{Zhang:2017bzy}, with very promising results for
$m_\pi=310 {\rm MeV}$, and with errors dominated by the uncertainty in
the Wilson-line self-energy. Nonetheless, the agreement with the largely
forgotten transverse-lattice calculations is worth
mentioning~\cite{Burkardt:2001mf,Dalley:2002nj}.

Along these lines Radyushkin has analyzed the  
nonperturbative $P_z$-evolution of parton quasidistributions~\cite{Radyushkin:2016hsy} and explored the connection between 
PDA, quasidistributions and the transverse-momentum distributions (TMD)~\cite{Radyushkin:2017gjd}.  
He also recalled the Ioffe-time distributions, where the longitudinal
distance rather than momentum display more clearly the physics of
the light cone kinematics~\cite{Braun:1994jq}, and introduced the
concept of pseudo-distributions~\cite{Radyushkin:2017cyf}, motivating
an exploratory study on the lattice~\cite{Orginos:2017kos}.

A key and subtle issue in the whole discussion regards the
renormalization. The meaning of renormalization in the transverse
lattice is quite transparent, as changing the renormalization scale $\mu$
corresponds to varying the transverse lattice spacing $a_\perp \sim
1/\mu$, such that for wavelengths smaller than $a_{\perp}$ all the dynamics
gets frozen and evolution stops. The original transverse lattice calculation of both PDF
and PDA of the pion~\cite{Dalley:2002nj} considered $a_{\perp} \sim 0.4~{\rm fm}$, 
corresponding to a low scale $\mu \sim 500~{\rm MeV}$, for
which the partonic expansion should be expected to converge.

The nonperturbative renormalization of nonlocal quark bilinears,
characteristic of parton quasidistributions, has been addressed by
several groups and, inspired by an analogy and experience with heavy
quark physics, it has  been recently found how the use of an auxiliary
spinless color field reduces the problem to the well understood
renormalization of local and composite operators within an effective
field theory viewpoint~\cite{Ji:2017oey}, as well as on the
lattice~\cite{Green:2017xeu}. Amazingly, the multiplicative
renormalizability of QDFs in coordinate space has also been
established to all orders in perturbation
theory~\cite{Ishikawa:2017faj}.
 
The purpose of this paper is to illustrate the above-discussed
concepts in an explicit nonperturbative model calculation for the
pion, where QDA and QDF for the valence quarks can be obtained
analytically. In our study, we use the Nambu--Jona Lasinio (NJL)
model~\cite{Nambu:1961tp,Nambu:1961fr} with quarks in the
Pauli-Villars (PV) regularization, as well as the Spectral Quark Model
(SQM)~\cite{RuizArriola:2001rr,RuizArriola:2003bs,RuizArriola:2003wi}.
In both models, all the theoretical constraints of chiral symmetry,
relativity, gauge invariance, and positivity are fulfilled for
PDFs~\cite{Davidson:1994uv,Davidson:2001cc,Weigel:1999pc},
PDAs~\cite{RuizArriola:2002bp} (for a review of chiral quark models
within the PDF and PDA context, see, e.g.,
Ref.~\cite{RuizArriola:2002wr} and references therein.), and
GPD~\cite{Broniowski:2007si}\footnote{In a model where a finite
cut-off must be introduced, the implementation of these constraints
without violation of positivity is highly non-trivial (see, e.g., a
recent discussion in Ref.~\cite{Tiburzi:2017brq})}.

We obtain very simple analytic expressions for QDA, QDF, their
transverse-momentum unintegrated analogues, as well as for the
Ioffe-time distributions. We illustrate with our formulas the
Radyushkin relation~\cite{Radyushkin:2016hsy}, linking via simple integration the QDA and QDF
to the transverse-momentum distributions corresponding to PDA and PDF.
Finally, we compare the QDA and PDA to the available lattice data and
find very reasonable agreement.

A major issue in nonperturbative model calculations is the proper
identification of the renormalization scale where the model is
supposed to work. We assume that in the chiral quark model, where no
other degrees of freedom are present, the valence quarks carry $100\%$
of the pion momentum. This renormalization condition sets the initial
scale for the QCD evolution, while phenomenological studies of parton
distributions for the pion extracted from Drell-Yan and prompt photon
experiments~\cite{Sutton:1991ay} yield $47(2)\%$ at $Q=2~{\rm GeV}$
for the valence quarks (see also
Ref.~\cite{Gluck:1991ey,Gluck:1999xe}), which is used to determine the evolution ratio. The perturbative scheme using
standard DGLAP to LO and NLO with $\mu \sim Q$, despite providing an
unusually low initial scale $\mu_0 = Q_0 = 313_{-10}^{+20}~{\rm MeV}$, 
gives small differences between LO and
NLO~\cite{Davidson:1994uv,Davidson:2001cc}.  As explicitly shown in
Ref.~\cite{Broniowski:2007si}, the approach mimics quite reasonably
the nonperturbative evolution observed in transverse lattice
simulations~\cite{Dalley:2002nj}.  It also predicts somewhat less
gluon content than
expected~\cite{Sutton:1991ay,Gluck:1991ey,Gluck:1999xe}, a problem
which has never been understood within the NJL model, as it requires
some gluon content present at the low scale $Q_0$, with the radiative generation of
gluons not sufficient.

For completeness, we mention lattice simulations for
the nucleon~\cite{Alexandrou:2015rja}, where a complete
non-perturbative renormalization prescription QDFs has been
carried out~\cite{Alexandrou:2017huk}.

Other model calculations of QDFs have also been performed.  A
quark-diquark model for the nucleon was
proposed in Ref.~\cite{Gamberg:2014zwa}.  More recently, QDA in a nonlocal
model for the pion was calculated~\cite{Nam:2017gzm}. However, the nonlocal model suffers from 
ambiguities related to the identification of the momentum fraction carried by the valence quarks,
and hence to the scale.

\section{Derivation}

In the NJL model (see e.g. Ref.~\cite{RuizArriola:2002wr} for a
review) sufficiently strong point-like 4-quark interactions lead to
dynamical chiral symmetry breaking, attributing the quarks, via a gap
mechanism, with a constituent mass $M \sim 300$~MeV.  The model is
suited for evaluation of soft matrix elements with pions (and
photons), in particular those relevant for PDA and PDF.

The model needs to be regularized to get rid of the ultraviolet
divergences, leaving the soft-momentum degrees of freedom in the
dynamics. The simplest regularization scheme satisfying the necessary
requirements is PV
regularization~\cite{Schuren:1991sc}, where the one-quark-loop
quantity $A(M^2)$ is replaced with
\begin{eqnarray}
&& \hspace{-1.4cm} A(M^2)|_{\rm reg} \equiv A(M^2) - A(M^2\!+\!\Lambda^2) +  \Lambda^2 \frac{d}{d\Lambda^2}A(M^2\!+\!\Lambda^2). 
\nonumber \\ && \hspace{-4cm} 
\end{eqnarray}
The PV cutoff $\Lambda$ is adjusted to reproduce the value of the pion
decay constant $f$. For $M=300$~MeV we use $\Lambda=731$~MeV in the
chiral limit, yielding $f=86$~MeV.  An interesting alternative for
regularization is offered by the Spectral Quark Model (SQM), where
one-quark loop is regularized by introducing a generalized spectral
density $\rho(w)$ of the spectral mass $w$, integrated over a suitably
chosen complex contour~\cite{RuizArriola:2003bs}. The construction
implements exact vector-meson dominance of the pion electromagnetic
form factor, namely $F(Q^2) = 1/(1+ Q^2/m_\rho^2)$, with $m_\rho^2 = 24
\pi^2 f^2 /N_c $ yielding $m_\rho= 764~{\rm MeV}$ for $f=86$~MeV.

The definition of the pion QDA is given by the matrix elements of the
bilocal quark operators,
\begin{eqnarray}
&& \hspace{-1.3cm}\tilde \phi(y,P_z) = \frac{i}{f} \int \frac{dz}{2\pi} e^{-i(y-1) z P_z} \times \nonumber \\
&& \langle \pi(P) | \overline{\Psi}(0) \gamma^3 \gamma_5 \Gamma(0,z) \Psi(z) | 0 \rangle, 
\end{eqnarray}
where $z$ is the spatial separation, $P=(E,0,0,P_z)$ denotes the four-momentum of the pion
(isospin indices are suppressed for brevity), 
and $y$ is the  fraction of $P_z$
carried by the valence quark, which now may assume any real value. The gauge link operator 
$\Gamma(0,z)$ is neglected in chiral quark models. 
The definition of QDF is analogous, but with the vacuum state replaced with $|\pi(P) \rangle$. The corresponding definitions of PDA and PDF 
involve $z$ constrained to the direction of the light cone, $z^+$. 

The evaluation of these distributions in the momentum space yields the
scalar two-point function (conventionally, all our momenta are
Euclidean)
\begin{eqnarray}
&& \hspace{-14mm}  I(x,P\cdot n,P^2,n^2)= \frac{4N_c M^2}{f^2} \int \frac{d^4k}{(2\pi)^4} \, \delta(x-k\cdot n) S_k S_{k-P}, 
\nonumber \\  && ~\hspace{-3cm} \label{eq:bubble}
\end{eqnarray}
where $S_k=1/(k^2+M^2)$, 
%$f$ is the pion decay constant, -- was earlier
$M$ is the
constituent quark mass, $P^2=-m_\pi^2$, and $n$ is the four-vector
yielding the kinematic constraint. For the PDA case it is a null
vector, $n^2=0$, whereas for QDA $n=(0,0,0,1/P_z)$, hence
$n^2=-1/P_z^2$. In both cases $P\cdot n=1$.  Using the Schwinger
$\alpha$ representation $S_k=\int_0^\infty d\alpha \exp[-\alpha
(k^2+M^2)]$ and proceeding along the lines of Appendix~A of
Ref.~\cite{Broniowski:2007si}, we can easily find the generic formulas
for PDA and QDA.  The scalar two-point function is
\begin{eqnarray} 
&& \hspace{-1.3cm} I =  \frac{N_c M^2}{4\pi^2 f^2} \int_0^\infty \!\!\!\! dk_T^2 \int_0^\infty \!\!\!\!\! d\alpha \int_0^\infty \!\!\!\!\! d\beta 
\int_{-\infty}^\infty \!\! \frac{d\lambda'}{2\pi} \times \label{eq:alpha} \\ 
&& \hspace{-.2cm} e^{-(\alpha+\beta)(k_T^2+M^2-\lambda'^2 n^2/4) +m_\pi^2 \alpha \beta/(\alpha+\beta) + i \lambda'  [\beta-(\alpha+\beta)x]}. \nonumber
\end{eqnarray}
In the PDA case of $n^2=0$, the integration over $\lambda'$ yields the distribution $\delta[\beta-(\alpha+\beta)x]$, which in turn provides the support $x \in [0,1]$, as 
$\alpha, \beta \ge 0$. However, when $n^2=-1/P_z^2 <0$, the  $\lambda'$ integration produces the function
$P_z \exp \{-(\alpha+\beta) [x-\beta/(\alpha+\beta)]^2 P_z^2 \} /\sqrt{\pi(\alpha+\beta)}$, leaving the support of $x$ unconstrained. 
In this case, following the established convention, we denote $x$ with $y$, which is interpreted as the fraction of the  longitudinal momentum 
of the pion carried by the quark (as already mentioned, 
it can be larger than 1 or negative, in which case the total momentum is compensated with the antiquark). 

Our objects of interest can now be readily identified with the
previous calculation. First is the QDA, equal to
\begin{eqnarray}
&& \hspace{-1.3cm} \tilde \phi(y,P_z) =   \frac{N_c M^2}{4\pi^2 f^2} \int_0^\infty \!\!\!\!\! d\alpha \int_0^\infty \!\!\!\!\! d\beta  \frac{P_z}{\sqrt{\pi}(\alpha+\beta)^{3/2}}\times
\label{eq:phialpha}\\
&& \hspace{-.2cm} e^{-(\alpha+\beta) \{M^2+ [y-\beta/(\alpha+\beta)]^2 P_z^2\} +m_\pi^2 \alpha \beta/(\alpha+\beta) } \bigg |_{\rm reg.}, \nonumber 
\end{eqnarray}
with explicit formulas for the two regularization methods,  following from the analytic integration over $\alpha$ and $\beta$, provided below. 
The other quantity of interest is the transverse-momentum distribution amplitude (TMA), obtained with Eq.~(\ref{eq:alpha}) with $n^2=0$ by leaving the $k_T$ momentum 
unintegrated. We find 
\begin{eqnarray}
&& \hspace{-1.3cm}  {\cal F}(x,k_T^2)= \frac{N_c M^2}{4\pi^2 f^2} \int_0^\infty \!\!\!\!\! d\alpha \int_0^\infty \!\!\!\!\! d\beta \delta[\beta-(\alpha+\beta)x]  \times 
\label{eq:TMAalpha}\\
&& \hspace{-.2cm} e^{-(\alpha+\beta) \{M^2+ k_T^2 \} +m_\pi^2 \alpha \beta/(\alpha+\beta) } \bigg |_{\rm reg.} \nonumber \\ 
&& \hspace{-.2cm} =   \frac{N_c M^2}{4\pi^2 f^2} \frac{\theta[x(1-x)]}{k_T^2+M^2 - m_\pi^2 x(1-x)} \bigg |_{\rm reg.}, \nonumber 
\end{eqnarray}
where $\theta$ is the Heaviside step function. 

With the formulas (\ref{eq:phialpha}) and (\ref{eq:TMAalpha}) it is
immediate to verify Radyushkin's formula (2.29) from
Ref.~\cite{Radyushkin:2016hsy}, namely
\begin{eqnarray}
\tilde \phi(y,P_z) = \int_{-\infty}^{\infty} \!\!\!\! dk_1 \int_0^1 \!\!\! dx \, P_z  {\cal F}(x,k_1^2+(x-y)^2 P_z^2). 
\end{eqnarray}
Thus, in accordance to Ref.~\cite{Radyushkin:2016hsy}, QDA can be obtained from TMA via a straightforward 
double integration. The above discussion carries over analogously to the case of QDF and TMD.

\begin{figure*}[tb]
\begin{center}
\includegraphics[angle=0,width=0.35 \textwidth]{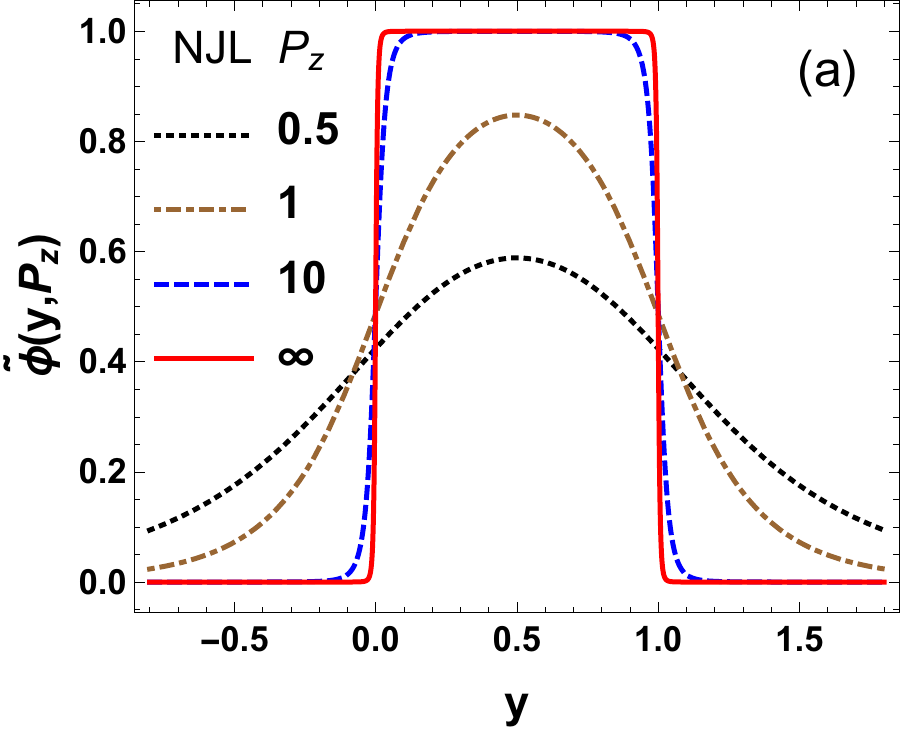} \hspace{.05\textwidth}  \includegraphics[angle=0,width=0.35 \textwidth]{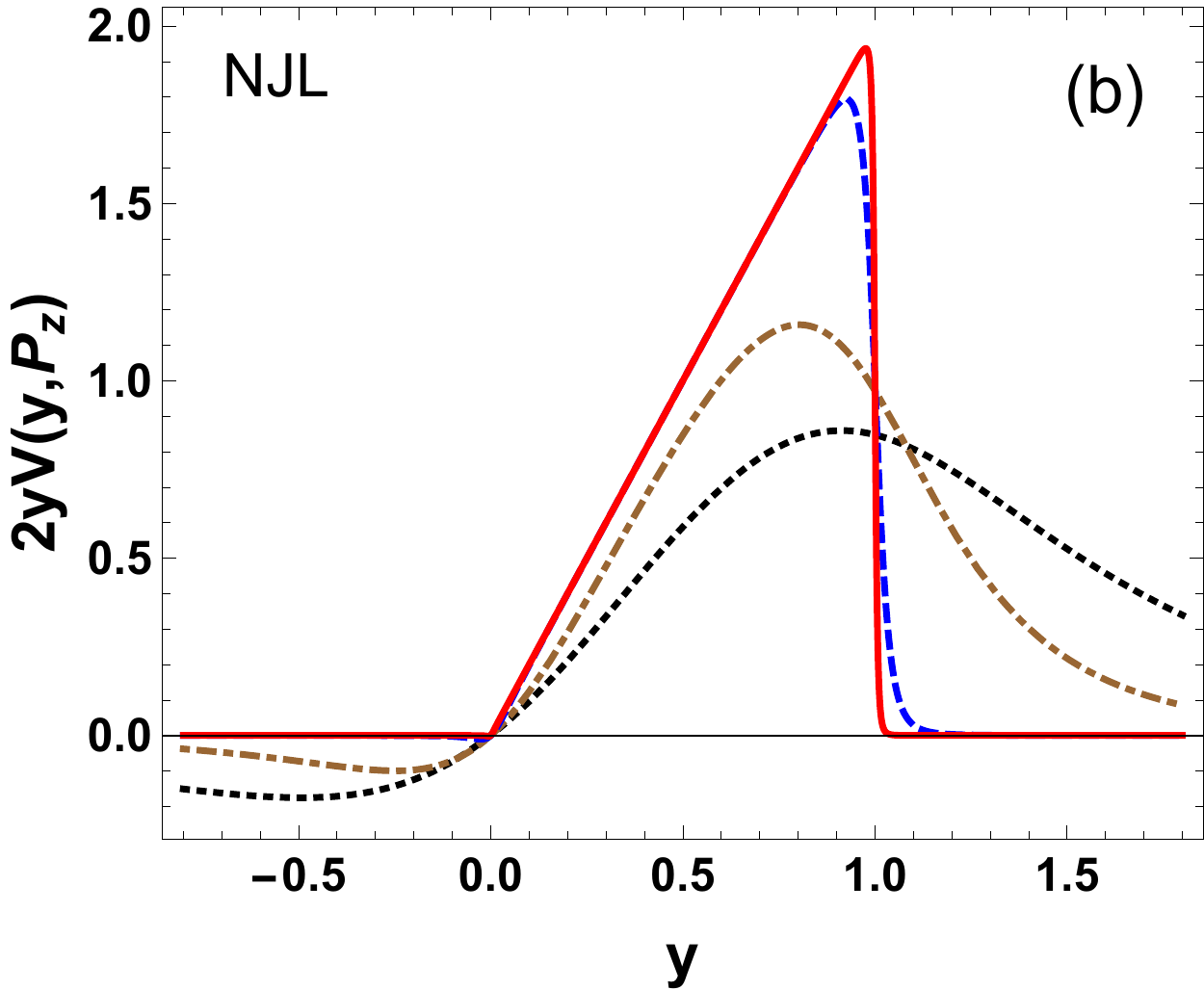}
\end{center}
\vspace{-6mm}
\caption{(a) Quark quasidistribution amplitude (QDA) of the pion in the NJL model, obtained at the constituent quark scale $\mu_0$ with $m_\pi=0$ at 
various values of the longitudinal momentum $P_z$, plotted as a function of the longitudinal momentum fraction $y$.
(b)~The same, but for the valence 
quark quasidistribution function (QDF) multiplied conventionally with $2y$. The legend indicates the values of $P_z$ in GeV. \label{fig:pz}  } 
\end{figure*}

The results for NJL and SQM are analytic, but they are particularly simple in the chiral limit of $m_\pi=0$, where we also have the feature that 
QDA is equal to QDF,  PDA to PDF, and TMA to TMD. In the NJL model, we have for the pion quasi wave function $\Psi$ (=TMA) and for the unintegrated 
valence quark quasidistribution function $V$ (=TMD) the expressions
\begin{eqnarray}
&& \hspace{-1.3cm}\Psi(y,k_T^2,P_z)=V(y,k_T^2,P_z) \label{eq:Psi} = \frac{N_c M^2}{8\pi^3 f^2} \times \\ && \hspace{-1cm}
 \left.  \frac{{P_z} y}{\left(M^2+k_T^2\right) \sqrt{M^2+k_T^2+P_z^2 y^2}} \right|_{\rm reg.} + (y\leftrightarrow 1-y), \nonumber
\end{eqnarray}
where 
\begin{eqnarray}
f^2= -\frac{N_c M^2}{4 \pi^2}\left .  \ln (M^2) \right |_{\rm reg.}.
\end{eqnarray}
The corresponding QDA and QDF for the valence quarks, obtained from Eq.~(\ref{eq:Psi}) via integration over $dk_T^2$, are
\begin{eqnarray}
&& \hspace{-1.3cm}\tilde \phi(y,P_z)=V(y,P_z) \label{eq:QDA} = \frac{N_c M^2}{4\pi^2 f^2} \times \label{eq:QDANJL} \\  && \hspace{-1cm} 
 \left . {\rm sgn}(y) \ln \left( \frac{P_z |y| +\sqrt{M^2+P_z^2 y^2}}{M} \right) \right|_{\rm reg.} + (y\leftrightarrow 1-y) , \nonumber
\end{eqnarray}
In SQM, the analogous formulas read:
\begin{eqnarray}
&& \hspace{-1.3cm}\Psi(y,k_T^2,P_z)=V(y,k_T^2,P_z) = \frac{m_\rho^3}{\pi^2 } \times \label{eq:PsiSQM} \\ && \hspace{-1cm}
\Bigg \{ \frac{3 \pi }{\left(4 k_T^2+m_\rho^2\right){}^{5/2}} 
-\frac{6 \, {\rm arcctg} ^{-1}
\left(\frac{2 P_z \sqrt{4 k_T^2+m_\rho^2}}{4 k_T^2+m_\rho^2+4 (y-1) y P_z^2}\right)}{\left(4 k_T^2+m_\rho^2\right){}^{5/2}}\nonumber \\ &&
\hspace{-1.33cm}
 + \frac{4P_z}{(4 k_T^2+m_\rho^2)} % \times \nonumber \\ &&  
 \left [ \frac{y \left(20 k_T^2\!+\!5 m_\rho^2\!+\!12 y^2 P_z^2\right)}{\left(4 k_T^2+m_\rho^2+4 y^2 P_z^2\right){}^2}
\!+\! (y \leftrightarrow 1\!-\!y )\right ] \! \Bigg  \}  \nonumber
\end{eqnarray}
and
\begin{eqnarray}
&& \hspace{-1.3cm}\tilde \phi(y,P_z)=V(y,P_z) \label{eq:QDASQM} = \\  && 
\frac{1}{2}-\frac{1}{\pi} {\rm arcctg}\left(\frac{2 m_\rho P_z}{m_\rho^2-4 (1-y) y P_z^2}\right) + \nonumber \\ && 
\frac{2 m_\rho P_z \left(m_\rho^2+4 (1-y) y P_z^2\right)}{\pi  \left(m_\rho^2+4 (1-y)^2 P_z^2\right) \left(m_\rho^2+4 y^2 P_z^2\right)}. \nonumber
\end{eqnarray}
The above quantities satisfy the proper normalization 
\begin{eqnarray}
&& \int_{-\infty}^\infty dy \,\tilde \phi(y,P_z)=\int_{-\infty}^\infty dy \, V(y,P_z) =1, \label{eq:norm}\\
&& \int_{-\infty}^\infty dy \,2y V(y,P_z) =1.  \nonumber 
\end{eqnarray}
and the limit
\begin{eqnarray}
\lim_{P_z \to \infty} \tilde \phi(y,P_z) = \lim_{P_z \to \infty} V(y,P_z) = \theta[y(1-y)].
\end{eqnarray}

We also obtain, in the chiral limit, very simple expressions for the
Ioffe-time distribution~\cite{Braun:1994jq,Radyushkin:2017cyf},
\begin{eqnarray}
{\cal M}(\nu, z_3) = \int_{-\infty}^\infty \!\! dy \, e^{i (y-{1\over 2}) \nu} \tilde \phi(y,P_z) \label{eq:Ioffe}
\end{eqnarray}
(we shift $y$ by ${1 \over 2}$ to get real expressions), where
$z_3=\nu/P_z$. The distribution (\ref{eq:Ioffe}) is normalized to ${\cal M}(0, 0)=1$, corresponding to Eq.~(\ref{eq:norm}). In the NJL
model we get 
\begin{eqnarray}
{\cal M}(\nu, z_3)= \frac{N_c M^2}{2\pi^3 f^2} \frac{\sin \left(\frac{\nu }{2}\right)}{\nu } K_0(M z_3) \bigg |_{\rm reg.},
\end{eqnarray}
whereas in SQM the result reads
\begin{eqnarray}
{\cal M}(\nu, z_3)= \frac{\sin \left(\frac{\nu }{2}\right) }{\nu} e^{-\frac{ m_\rho z_3}{2}} \left(m_\rho z_3+ 2 \right).
\end{eqnarray}
We note the dependence on $\nu$ and $z_3$ is factorized. The
large-$z_3$ behavior of the two models is similar, as $K_0(t)\sim
e^{-t}/\sqrt{t}$, and $m_\rho \simeq 2M$.

\section{Numerical Results}

After listing in the previous section simple formulas obtained in the
chiral quark model for QDA of the pion and the related quantities, we
now show our results numerically and compare to the available lattice
data~\cite{Zhang:2017bzy}. Note that whereas QDA and QDF have
originally been thought of as auxiliary quantities needed to obtain
PDA and PDF on Euclidean lattices, they may well be used to verify model
predictions obtained at particular finite values of $P_z$, thus acquiring 
practical significance in their own right.

In Fig.~\ref{fig:pz} we present the quark quasidistribution amplitude
(QDA) and the valence quark quasidistribution function (QDF) of the
pion in the NJL model, obtained at the constituent quark scale $\mu_0$
and in the chiral limit of $m_\pi=0$, evaluated according to
Eq.~(\ref{eq:QDANJL}). The results are given for various values of the
longitudinal pion momentum $P_z$ and are plotted as functions of the
fraction of $P_z$ carried by the valence quark, $y$. We note that as $P_z$
increases to infinity, the curves tend to the PDA and PDF at the quark model scale
$\mu_0$, which are equal to $\phi(x) = \theta[x(1-x)]$ and $2x V(x) =
2 x \theta[x(1-x)]$, respectively. Evolution in $\mu$ is certainly necessary for PDA
and PDF to compare these quantities to the data obtained at higher scales $\mu$
(cf.~Fig.~\ref{fig:DA} in the following and its discussion).

\begin{figure*}[tb]
\begin{center}
\includegraphics[angle=0,width=0.35\textwidth]{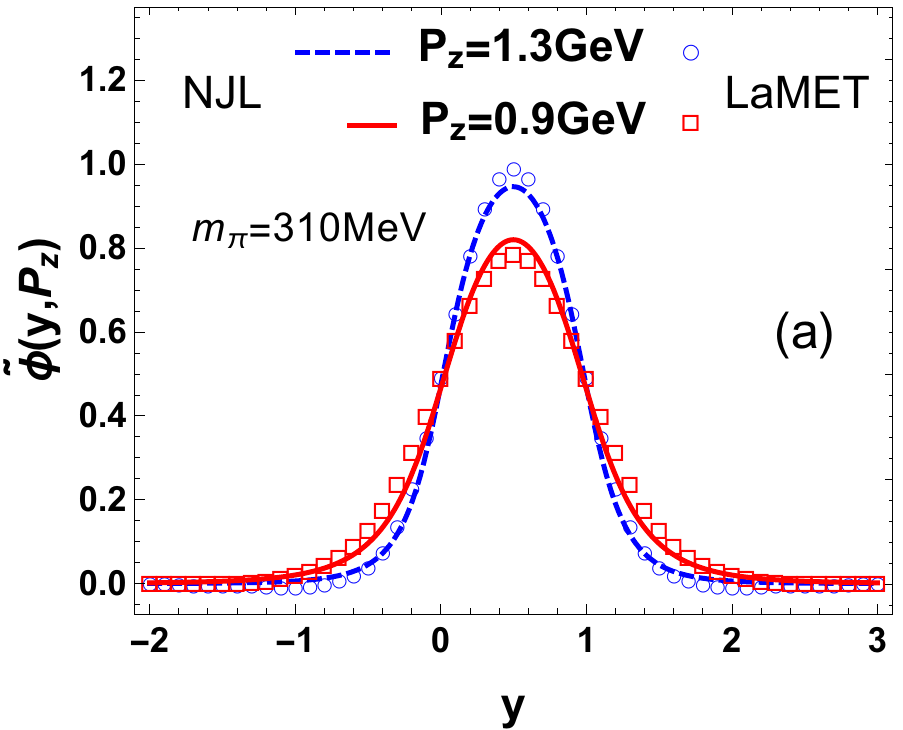}  
\hspace{.05\textwidth}  \includegraphics[angle=0,width=0.35\textwidth]{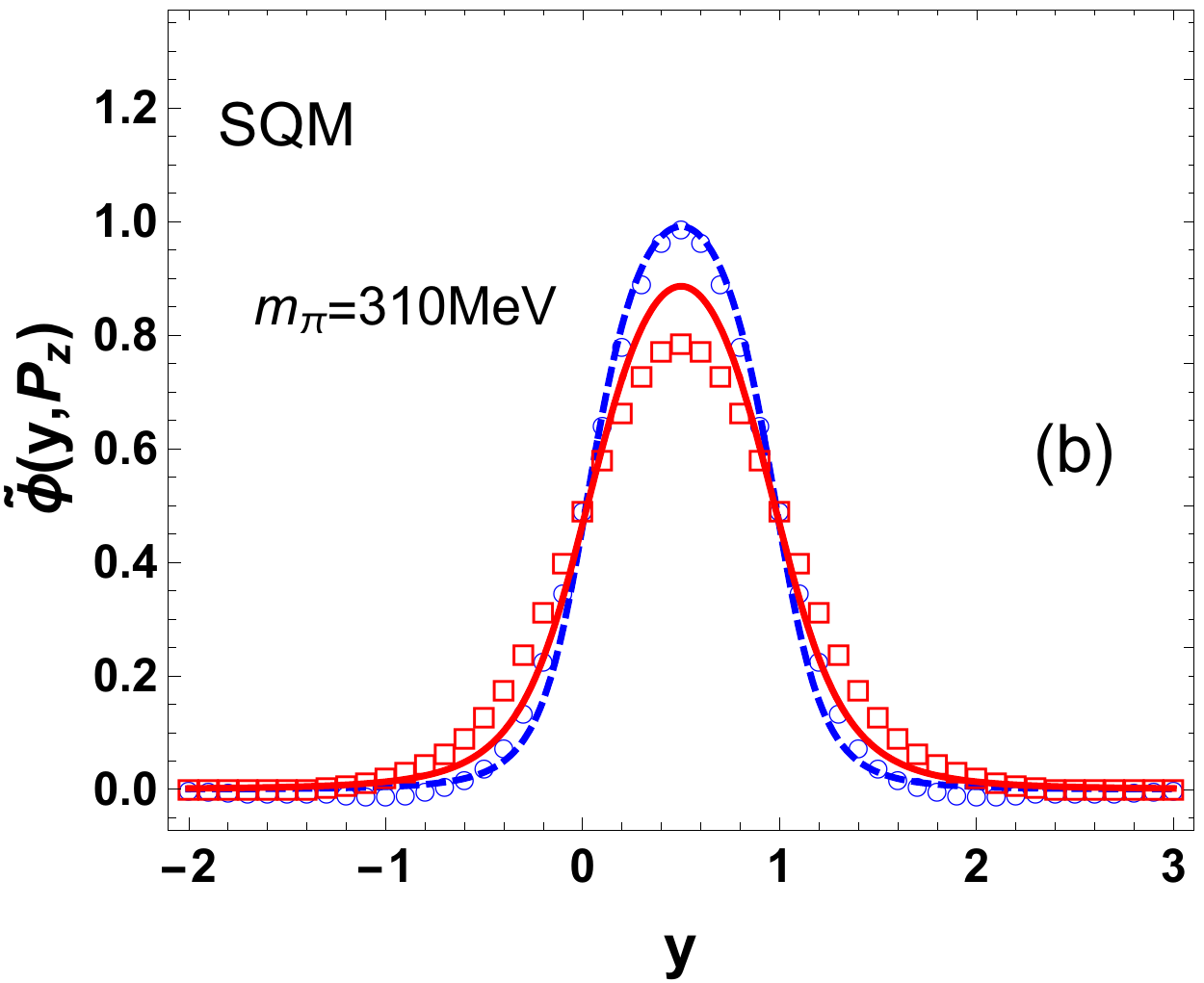}
\end{center}
\vspace{-6mm}
\caption{Quark quasidistribution amplitude (QDA) of the pion in the NJL (a) and SQM (b) models at the constituent quark scale $\mu_0$ and with $m_\pi=310$~MeV,
plotted as functions of the longitudinal momentum fraction $y$, evaluated for $P_z=0.9$ and $1.3$~GeV, and
compared to the lattice data at $\mu=2$~GeV from the LaMET Collaboration, used in Ref.~\cite{Zhang:2017bzy}. \label{fig:lat} }
\end{figure*} 

We compare our NJL model~(a) and SQM~(b) results for QDA of the pion to the lattice data in Fig.~\ref{fig:lat}. The model 
curves are obtained at the constituent quark scale $\mu_0$, whereas the lattice data correspond to the scale $\mu=2$~GeV, as inferred from the lattice spacing. 
The presented points were used in the extrapolation of the LaMET Collaboration data in 
Ref.~\cite{Zhang:2017bzy} to obtain the pion PDA. 
We use $P_z=0.9$ and $1.3$~GeV and $m_\pi=310$~MeV,  exactly as used in  Ref.~\cite{Zhang:2017bzy}.

While the agreement of a direct comparison is quite remarkable, one
should note that there is room for improvement, particularly from the
QCD evolution point of view. In a future work, the model QDA could be
evolved in $\mu$ with the equations outlined in Ref.~\cite{Ji:2015qla}
and modified for the lattice calculations~\cite{Zhang:2017bzy} to
account for a finite cut-off effect with a Wilson line self-energy 
correction~\cite{Chen:2016fxx}.

For PDA or PDF, the evolution equations have been efficiently
implemented (ERBL~\cite{Efremov:1979qk,Lepage:1979zb} and
DGLAP~\cite{Gribov:1972ri,Altarelli:1977zs}) and used to
evolve the model results from $\mu_0$ to experimental or lattice
scales~\cite{Davidson:1994uv}.  The results of this evolution, from
the quark model scale $\mu_0=313$~MeV, where PDA of the pion is constant in the
chiral limit, $\phi(x)=\theta[x(1-x)]$~\cite{RuizArriola:2002bp}, to
the scales $\mu=2$~GeV and $\mu=0.5$~GeV are given in
Fig.~\ref{fig:DA}, where we also compare to the Euclidean lattice
extraction~\cite{Chen:2017mzz} and to the transverse lattice
data~\cite{Dalley:2002nj}.

\begin{figure*}[tb]
\begin{center}
\includegraphics[angle=0,width=0.34 \textwidth]{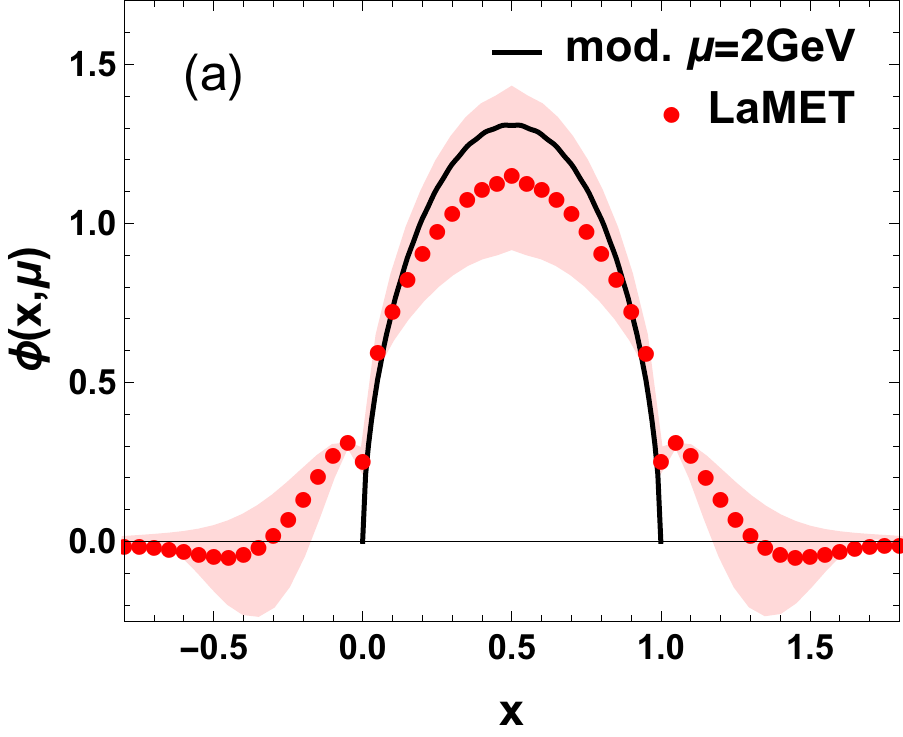}  \hspace{.05\textwidth}  \includegraphics[angle=0,width=0.34 \textwidth]{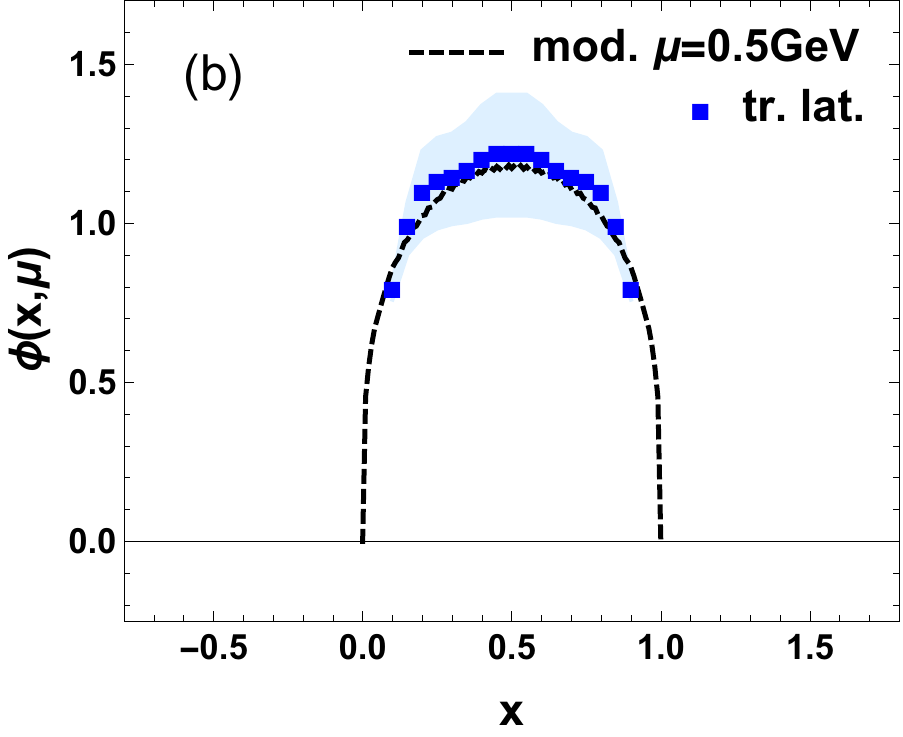}
\end{center}
\vspace{-7mm}
\caption{The distribution amplitude (PDA) of the pion, obtained from the chiral quark models (the same for NJL and SQM) with $m_\pi=0$, evolved to the scale 
$\mu=2$~GeV~(a) and $\mu=0.5$~GeV~(b), and compared to the lattice data of the LaMET Collaboration~\cite{Chen:2017mzz}~(a) 
and to the transverse-lattice results~\cite{Dalley:2002nj}~(b). \label{fig:DA}} 
\end{figure*} 

We recall that some low moments of the parton distributions of the pion on
the Euclidean lattice have been evaluated. In the case of PDF, the
second moment corresponding to the pion momentum fraction at $\mu = 2$ GeV
in the $\overline{\rm MS}$-scheme, takes the value $\langle x \rangle_{\pi}^q
=0.214(15)_{-9}^{+12} $~\cite{Abdel-Rehim:2015owa}. In the case of 
PDA, only the second moment has been extracted and the most accurate
result, also at $\mu= 2$~GeV in the $\overline{\rm MS}$-scheme, is found
to be $\langle \xi^2 \rangle = 0.2077(43)(32)$, or equivalently, 
$ a_2 =0.0762(127)$~\cite{Bali:2017ude} for the second Gegenbauer coefficient. 
In the chiral quark models evolved from
$\mu_0=290$~MeV to $\mu= 2$~GeV, we get $\langle x \rangle_\pi^q
=0.21$ and $a_2=0.10$, which is compatible with the above-quoted
lattice values.

We remark that one of the features implied by our choice of the
normalization scale $\mu \sim Q$ and the low energy scale $\mu_0$ is
the large-$x$ behavior of the PDFs, where {\em after} the DGLAP
evolution one has
\begin{eqnarray}
\hspace{-0.7cm} V(x,Q)/V(x,Q_0) \sim (1-x)^{-(C_F /2 \beta_0) \ln [\alpha(Q)/\alpha(Q_0)]}, 
\end{eqnarray}
which for $V(x,Q_0)=1$ yields $V(x,Q) \sim (1-x)^{1.1 \pm 0.1}$ for
$Q=2~{\rm GeV}$~\cite{RuizArriola:2002wr}, complying quite accurately
with experimental extractions~\cite{Conway:1989fs}. Actually, the
large-$x$ dependence of PDF of the pion has been a subject of
controversy~\cite{Wijesooriya:2005ir,Holt:2010vj} (see, e.g., Fig.~8 of
Ref.~\cite{Broniowski:2007si} for a comparison). From this viewpoint, a
QDF-based lattice calculation would be most helpful to settle the
issue.

\section{Conclusions} 
 
Hadronic physics on the light cone has been a topic of
phenomenological and experimental discussion ever since the parton
model was proposed by Feynman, but the actual {\it ab initio}
calculations of the corresponding parton distributions confront
enormous theoretical and practical difficulties.  In this paper, we
have analyzed, within chiral quark models of the pion, the recently
proposed quasi-distributions. These objects have been proposed as
means to extract, from Euclidean lattices, the space-like correlators
boosted to the infinite momentum frame.

Our model calculation describes the pion as a composite $q \bar q$
state and complies with all the {\it a priori} requirements imposed by the chiral symmetry
the gauge invariance, relativity, and positivity; it
provides a helpful and convenient playground to explore the peculiar
features of the new way of extracting the Minkowskian parton distributions from the
Euclidean formulation.  We believe that our study is useful from two
viewpoints. First, it provides nonperturbative predictions for the
soft objects of interest from a dynamical model, with the results
expressed by very simple and intuitive formulas.  The predictions for
the quasidistributions may be confronted with the lattice
extractions, whereby these quantities acquire practical meaning. 
On the other hand, our formulas may be used to illustrate
and verify the methods and various relations between the pertinent quantities, such as for
instance the Radyushkin relation of the quasidistributions with the
transverse-momentum distributions.

We have shown that our model (without the QCD evolution implementing radiative corrections) 
compares favorably to the recent lattice
extractions of the valence quark PDA of the pion, and thus mimics quite
accurately the space-like boosted dynamics observed on the lattice.  We also
find that finite pion mass effects are a few percent away from the
chiral limit. However, the finite pion mass destroys the factorization
between the transverse momentum and the Bjorken $x$ variable.

We also predict similar features for the QDF and PDF of the pion, suggesting
that a lattice calculation based on quasi-distributions should be
equally feasible and realistic, as it has been in the PDA case. Therefore, most
interesting would be an analysis of the large-$x$ behavior and the
extension to singlet distributions, as the chiral quark models fail to fully reproduce the
phenomenologically needed gluon content of hadrons.  We hope that our paper
will stimulate such studies on the lattice, hence filling the gap in
our understanding of the pion structure on the light-cone from a fundamental QCD viewpoint.

%\bigskip

We thank Jianhui Zhang for kindly providing the data from Ref.~\cite{Zhang:2017bzy}, including 
the unpublished points for the QDA shown in our Fig.~\ref{fig:lat}.
This work was supported by the Spanish Mineco (Grant FIS2014-59386-P), the
Junta de Andaluc\'{\i}a (grant FQM225-05), and by the Polish National
Science Center grant 2015/17/B/ST2/01838.

\vspace{-2mm}

\bibliography{quasiNJL}

\end{document}